# Report on Data Quality in Biobanks: Problems, Issues, State-of-the-Art[1]

Suneth Ranasinghe, Horst Pichler, Johann Eder

Alpen-Adria-Universität Klagenfurt

Abstract

This report discusses the issues of data quality in biobanks. It presents the state-of-the-art in data quality: the definition of data quality, the dimensions of data quality, and the quality management system for achieving or describing the aspired data quality characteristics and we present and discuss all elements for such a data quality management system.

In depth, we discuss the requirements and the context of data quality for biobanks in particular, where we argue that biobanks can be seen as data brokers, where the indented use of the data is to support the search for suitable material and data in the preparation of medical studies. For such an intended use, the quality of the metadata is of high importance and biobanks have to emphasize to strive for adequate documentation of the quality of the data annotating the samples.

[1] The work on this paper has been supported by bmwfw - the Austrian Ministry for science, research and economics within the project bbmri.at (GZ 10.470/0016-II/3/2013)

# Table of Contents

# 1. Introduction

The quality of data plays an important role in the biomedical research domain. Insufficient reproducibility of data from medical studies have recently been discussed together with the highly problematic consequences for the advancement of science, the economics of research and the development of suitable drugs and therapeutic procedures (see e.g. [3,16,19]). The research results highly depend on the quality of the data, which finally helps to form the conclusions. The absence of clear and uniform procedures and standards as well as unawareness of the quality of material and data used for performing studies may be dangerous, when operationalizing research. Further, the levels of skill, ability and knowledge not only differ from one person to another, but may even differ in the same person depending on circumstances (e.g., fatigue can degrade the performance of a skilled operator). Thus, it is important to know about the quality of data and define proper procedures and standards to achieve it in the same way, as it is important to develop standard operating procedures implementing prevailing quality standards for dealing with the biological material (samples) stored in the biobanks. The intention of this report is to provide information about the importance of data quality, data quality dimensions, how to achieve data quality, and data quality measurements and assessments central to bio medical research.

## 1.1. What is data?

Data is an abstract representation of selected characteristics of real-world objects, events, and concepts, expressed and understood through explicitly definable conventions related to their meaning, collection, and storage. Each piece is important as abstract representations about data though it may not be the "reality" itself. It could be a selected characteristic of real-world objects, events, and concepts but may not reflect every characteristic about the object. Data is expressed through explicitly definable conventions related to their meaning, collection, and storage that defined in ways that encodes meaning. Choices about how to encode are influenced by the ways that data is created, used, stored, and accessed.

The New Oxford American Dictionary (NOAD) defines data as "facts and statistics collected together for reference or analysis." ASQ defines data as "A set of collected facts" and identifies two kinds of numerical data: "measured or variable data … and counted or attribute data" and ISO defines data as "re-interpretable representation of information in a formalized manner suitable for communication, interpretation, or processing" (ISO 11179).

On the bio medical perspective, bio medical databases store data, about patients and drugs, which are collected from humans, and machinery equipments, e.g. labs, X-ray. Such data are stored as textual data, or 2D data, e.g. image, X-ray, video data.

## 1.2. Data quality definition

The level of the data quality depends on the degree to which the data meets the expectations of the consumers, and their intended use of the data. Typically, the high-quality data meets the expectations of the users and exposes the representational effectiveness to a greater degree than low-quality data. Therefore, assessing the quality of data requires understanding those expectations and determining the degree to which the data meets them.

The Institute of Medicine (IOM) defines quality data as "data strong enough to support conclusions and interpretations equivalent to those derived from error-free data" [11]. Like Joseph Juran's famous "fitness for use" definition [20]. The purpose of a biobank is to collect, store and distribute high quality samples and data and it may, in addition, process and test the samples. The way in which the biobank performs these tasks needs to be controlled. So that all of the operations of the biobank, including the ways in which the biobank is managed and in which legal and ethical requirements are met, are "fit for purpose".

Typically, the organizations control the quality of their activities by implementing a quality management system (QMS). The QMS defines the organization's quality policy and objectives and ensures that these are achieved through quality assurance (QA) and quality control (QC) processes. QA focuses on the processes through which the product is obtained, whereas QC focuses on the product. Most scientists are familiar with QC, which is "that part of quality assurance that focuses on fulfilling quality requirements" and perform QC as a routine part of their day-to-day activities. QC consists of specific tests defined by the QA or QMS program to be performed to monitor procurement, processing, preservation and storage, specimen quality and test accuracy. These tests may include but are not limited to: performance evaluations, testing and controls that to determine the accuracy and reliability of the biobank's equipment and operational procedures as well as monitoring of the supplies, reagents, equipment and facilities.

## 1.3. Biobank perspective

Biobanks (see e.g. [29, 2, 12]) are collections of biological material together with data describing the source of the material, data about handling the material and data. The context and the specific characteristics of the production, collection and use of data in biobanks need a specific view on data quality and specific policies for defining and managing data quality. Typically the mainstream definition of DQ (fitness for use, etc.) is problematic for biobanks, as its intended usages are (frequently) not known in advance. Biobanks collect data and material (mostly by following a predefined procedure for quality assurance) and store it for a future use without being able to know the exact usage of the collected samples and the collected data. So the quality of the data should be documented properly from the beginning of its origin point. Also, the data provenance, i.e., how, by whom, and with which means had the data been produced/collected are very valuable information that are required to be continuously documented.

Many biobanks act as **"data brokers"** rather than data producers. Normally, the biobanks are responsible for storing the data collections and provide them for the researchers, whenever they request a particular sample from the biobank. Most of the data, which a biobank provides for its users, is, however, not produced by the biobank but is collected by the biobank from medical information systems, health records, laboratory information systems, pathology reports, etc. Typically, in hospital centered biobanks, the type of biobanks prevailing in Austria, the biobank itself mainly produces data about the harvesting and storing of the samples, but not the data with which the samples are annotated.

As the data stored in the biobank are extremely sensitive personal data, the importance of precautions to guarantee the privacy of the data owners (i.e. the donors of the biological material) has to be emphasized [13], but the quality of the data protection system is not subject of this report.

## 1.4. Distinction of data origin

Documenting the data origin information is therefore an important task, as it may be a valuable information for its use in research projects. Some of the possible examples of the data origin are as follows;

- Data that are produced by biobanks (BB), mostly the data about sample handling, such data attributes include information such as ischemia time, storage temperature, etc., that are useful information for the intended research work.
- Data that are explicitly produced for the biobank, for example when data is collected together with the samples for cohort studies.
- Data produced by scientific studies and collected in the biobank, e.g., data about the donors of samples collected in clinical departments in the course of research projects, data that are produced from establishing disease based collections.
- Data produced in routine health care, e.g., the continuous update about patients' personal data, lab reports, diagnosis, etc.
- Other meta data, describing the data.

Furthermore, it is necessary to document the data related information, i.e. meta-data, of the samples at a biobank. Some of the possible meta-data of the biobank collection data, are data about the source of the data items, protocols for collecting the data in the course of scientific studies, quality descriptions of the data such as precision, time of collections, method of collection, etc.

## 2. Quality Dimensions

Quality assurance is recognized as one of the crucial processes of biobanking collections and especially, it is required as these collections are shared and used by different organizations. Quality assurance needs to document the necessary context to interpret the data correctly. This requires biobanks to confirm that they follow the best practice guidelines and for achieving the defined high quality standards [10].

### 2.1. Data quality dimensions

Quality dimensions can be referred either to the extension of data—to data values, or to their intension—to their schema. Following are the data quality dimensions that are defined in [23], which are useful for preserving the quality of the data.

*Table 1: Data Quality (DQ) dimensions*

| DQ dimensions | References |
|---|---|
| Accuracy | [23], [6], [27], [4], [25] |
| Currency | [23], [6], [25], [18], [8] |
| Completeness | [23], [6], [25], [26], [18], [8], [24], [22] |
| Readability | [6] |
| Reliability | [6] |
| Usefulness | [6] |
| Cost-effectiveness | [6] |
| Confidentiality | [6] |
| Consistency | [23] |
| Timeliness | [23], [26], [8], [24], [22] |
| Relevance | [23] |

| Granularity | [23] |
|---|---|
| Specificity | [23] |
| Precision | [23] |
| Attribution | [23] |
| Volatility | [18], [8] |

According to the definition of each quality dimension, the data quality can be measured as high, medium or low. Brief explanation about each quality dimension is discussed using examples as follows;

### 2.1.1. Accuracy

Health data represent the truth about what actually happened to a particular person. Such data is important and should define the truth with respect to the data value itself. For example, does a heart rate of 92 represent the patient's true heart rate at the time of measurement? According to [23], accuracy is a state in the data match with the intended state in the real world. Also, [27] defined accuracy as "the extent to which data are correct, reliable and certified."

In general, data are accurate when the data values stored in the database correspond to the real-world values they represent [4]. It is a measure of the proximity of a data value, v, to some other value, v', that is considered correct [25]. For example, if in a data collection data is noisy, the available values are not precise or too vague, or obscured from the real world observations, then such data is considered as low quality data. Data which are low in quality, from the accuracy perspective, may lead to have more vague results than expected.

### 2.1.2. Currency

It is the measure of the data that is recorded at the time of the observation, and keeps track of whether it is up-to-date. Also, it can be considered as the time length of a data value that has been stored, since its last update [23]. Currency is the degree to which a data is up-to-date. A data value is up-to-date, if it is correct in spite of possible discrepancies caused by the time related changes to the correct value [25]. It can also be described as a measure of how old the information is, based on how long ago it was recorded [8].

If an available data collection is out dated or obsolete or may not be properly up-to-date, such collection is considered as low quality data collection in currency dimension perspective.

### 2.1.3. Completeness

Completeness is the dimension that measure the extent of the captured data attributes, with reflect to the real-world state of the data. It is the property of a set of data values; i.e., all the required data, that should be available to make an informed decision [23]. It is the degree to which values are included in a data collection [25]. According to [18], it is the percentage of real-world information entered in data sources and/or data warehouses. It is also described as the ratio between the number of non-null values in a source and the size of the universal relation [24]. [8] states the completeness as the information that have the all required parts of an entity's description. Also, [22] describes the completeness as all the values that are supposed to be collected as per a collection theory.

For example, consider a sample description data which consists of missing data attributes, missing values or the data that are partially available. Such data sample is considered as a low quality data sample.

### 2.1.4. Readability

It is the measure of describing that all the represented data are understandable. Such data could be represent as written, handwritten, transcribed, or printed [6]. If the data collection descriptions are not clear enough to read and to understand or if an image or video data is of low resolution, or noisy, it is consider as low quality data.

### 2.1.5. Reliability

Reliability indicates, how much you can trust that the data is a true representation of reality. It combines both the question whether the underlying concept is measured by suitable indicators and whether the assessment is done by qualified persons using suitable methods, e.g., is this question a reliable and valid measure of depressive mood? [6].
For an example, consider a data item "cause of death" collected by a general practitioner or by a trained pathologist after an autopsy the data collected by a trained pathologist is considered as more reliable, as he is more specialized doing such tasks than a general practitioner and an autopsy is a more accurate method.

### 2.1.6. Usefulness

According to [6], those are the only data that should be collected for care and secondary purposes. As mentioned in [6], usefulness is "for specified, explicit and legitimate purposes and not further processed in a way incompatible with those purposes". If there is no clear information specified about the collected data usages, such data is considered as low quality data in the perspective of the usefulness. This criterion of usefulness, however, is of less importance for biobanks, as biobanks collect data for future use, for studies with yet unknown requirements.

### 2.1.7. Cost-effectiveness

It is the data dimension that evaluate the cost of collecting and dissemination information, and typically this must not exceed its intended value purposes. If the process of collecting data is expensive, than it is dissemination information and its specified use, such data is considered as low quality data.

### 2.1.8. Confidentiality

Confidentiality is the aspect of the data availability for the authorized persons, when and where needed [6]. Once a data collection process started, it is important to know its intended data sharing methods, i.e., among which persons the data should be distributed or share with, to protect the data confidentiality. So it is quite necessary to know the data usage and it is sharing authorities from the initial stage of the data collection process. If it is not properly defined for a given data sample such data sample is considered as low quality data.

### 2.1.9. Consistency

It is the measure of data values compared to its real-world state. Such data representations should not be conflicted [23]. From the consistency perspective, if the data collection shows many contradictions and inconsistent data descriptions with the real-world state, such data is considered as low quality data.

### 2.1.10. Timeliness

It is the time length from a change in the real-world state to the time of the data that reflects the change [23]. Timeliness refers only to the delay between a change of a real world state and the resulting modification of the information system's state [26]. According to [8], it has two components: age and volatility. Also, [24] state timeliness as the average age of the data in a source and [22] state timeliness as the extent to which data are sufficiently up-to-date for a task [22].

If the data descriptions are not continuously up-to-date such data is considered as low quality data. For example, how recent that the body weight has been measured, is a critical data attribute when prescribing drugs for a patient. If it is not up-to-date, then the intended results or the prescribed drug details may be deviated.

### 2.1.11. Relevance

Relevance is the measure that is used to distinguish, whether the available data can be used to answer a particular question [23]. If the available data are irrelevant and not useful to be used for a specific task or to answer an intended question such data are considered as low quality data.

### 2.1.12. Granularity

It is the level of the detail that a particular data has been captured [23]. Knowing the level of the available data granularity is important, but if it is too fine grain or too coarse grain, then such data is low quality data.

### 2.1.13. Specificity

Specificity is the measure of each state in the data definition (metadata) corresponding to one (or no) state of the real world [23]. If the available data are too generic for an intended task such data are considered as low quality data.

### 2.1.14. Precision

Precision is the number of significant digits to which a continuous value was measured (and recorded); i.e., for categorical variables, the resolution of the categories [23]. If the collected data descriptions are not exact enough they might be useless for some studies. In the same way, if the precision of some data is not known, the data is less valuable for future uses. For an example, consider the data collected from a tentative diagnosis through fast test vs. a final diagnosis based on a thorough test. Typically, a fast test collects only a least amount of data that requires only to check the minimal potentiality of a diagnosis.

But diagnosis can be clearly identified only after an evaluation result of a thorough test, which provides data with a higher precision. Hence, such thorough test data typically is considered as high quality data.

### 2.1.15. Attribution

Source and individual generating and updating data are inextricably linked to data values [23]. It is important to know the exact origin of the data and the information about its distribution as the intended tasks and its future use may differ based on this information. So if such data attributes are not available for a particular data collection then such data is considered as low quality data.

### 2.1.16. Volatility

Volatility is a measure of information instability, the frequency of change of the value for an entity attribute [8]. [18] describes volatility as the time period for which information is valid in the real world.

It is quite necessary to capture such data volatility information continuously and be aware of the data changes in advance. In the perspective of volatility dimension, the unawareness of the frequent data change is an indication of a low quality data collection.

## 2.2. Important quality dimensions for biobanks

Though it seems that all the mentioned dimensions in section 4 are important to have a quality system, there are some dimensions, we consider in particular important for biobanks. Below we give some examples with corresponding quality dimensions.

- **Reliability**: Consider two data samples that are available about the death of a person that happened due to a stroke, which is reported by a general practitioner and by a trained pathologist. Concerning the quality of the data from the "reliability" perspective, data presented by the trained pathologist is considered as more reliable compared to that from the general practitioner.
- **Completeness**: We continue with the same example, about the "cause of death". Completeness is the percentage of available samples in the collection for which reliable data about the cause of death is available.
- **Accuracy**: Consider the accuracy of an immunity assumption due to titer analysis versus a patient remembering childhood diseases.
- **Precision:** Consider a data collection where data could be derived by from a quick test and a thorough test. Typically, a quick test is only carried out to exclude potential diagnosis, as quick tests are designed to have an extremely low number of false negatives even at the expense of higher numbers of false positives. Thorough tests, on the other hand are designed to reduce both false positives and false negatives at the expense of a longer and more costly procedure. To know the precision of such it is therefore necessary to know with which test these data were derived and which are the statistical error descriptions of these tests.
  Another example for precision is whether the ischemia time is derived from a statistical estimate according to the applied standard operating procedures or from case based measurements, i.e. whether the steps in each instance of the process are documented.
- **Timeliness:** Consider a body weight of a person. If a person undergoing a continuous medical treatment for a particular disease, a regular measure of body weight is important as it directly effects the medical treatments. If the treatments are based on outdated or obsolete data, then the provided treatments may not be accurate as the body weight data is not current at the time of the treatment.

## 2.3. Biobanks as data brokers

The main stream definition of data quality, i.e., fitness for use, is problematic for biobanks as the use of such data is not known (frequently) in advance. Hence, many biobanks act as data brokers rather than being actual data producers. The intended use is then the search for suitable samples. A major expectation of biobank users are the quality of the samples and the quality of related data documentations. Concerning the quality perspectives maintaining quality documentation is also an important fact. Thus for data brokers the primary target of data quality is to have a good quality of the meta-data, i.e. quality data documentation.

Table 2 shows the data definition levels of a biobank.

*Table 2: data definition levels*

| **Data** | Actual data associated with a sample |
|---|---|
| **Meta-data** | Data about data, in particular, data about the quality of data, i.e., data quality documentation. Frequently on the level of a collection where all samples and data have the same quality characteristics. |
| **Meta-metadata** | Data about Meta-Data, here: data about the quality of the data quality documentation. Frequently on the level of biobanks or on the level of collections. |

It is necessary to document the quality of the received data from the beginning of the process. It should include information such as data origin, data provenance, i.e., how, by whom and with which means had the data been produced/collected (e.g. quick test vs. diagnose test).

However, a major expectation of the biobank users is to search quality samples with properly maintained documentation. The intended use is then the search for suitable samples. Maintaining of high quality meta-data information, i.e., high quality documentation, is important, as it is the base reference of the data provenance. Hence, biobanks should assess and document the quality of the DQ documentation of samples and data items continuously.

## 2.4. Data Provenance

Data provenance is defined as the "description of the origins of a piece of data and the process by which it arrived in the database" [9]. Typically, the data annotations, e.g., comments about the data or other types of metadata, are used to trace the provenance of the data. Annotations can be used in a variety of situations including [7]:

1. Systematically trace the provenance and flow of data, namely, even if the data has undergone a complex process of transformation steps, we can determine the origins by examining the annotations.
2. Describe information about data that would otherwise have been lost in the database, e.g., an error report about a piece of data.
3. Enable the user to interpret the data semantics more accurately and to resolve potential conflicts among the data retrieved from different sources. This capability is useful in the field of data integration, where we are interested in understanding how data in different databases with heterogeneous semantics and different quality levels can be integrated.
4. Filter the data retrieved from a database according to quality requirements.
5. Improve the management of data trustworthiness through annotations referring to the reputation of a source or to a certification procedure.

## 2.5. Data quality documentation

DQ documentation is one of the best practice in DQ processes. It is always necessary to document all the related information about the data source.

Source documentation, i.e., bio sample meta-data, should follow the "ALCOA" attributes [5]:

### 2.5.1. Attributable

The data quality documents should have clearly documented the data, e.g. the link to the source, the information about the person who observed and recorded the information, and the relevant timestamps [17] should be clear and understandable.

### 2.5.2. Legible

The source documents should be readable and their signatures should be identifiable, i.e., whether the information provided are easy to understand, whether the data are recorded permanently on a durable medium, and whether the original entries have been preserved, etc. [17].

### 2.5.3. Contemporaneous

The related information should be documented in the correct time frame along with the flow of events. If a clinical observation cannot be entered when it is made, at least the chronology information should be recorded. However, if there is a delay, it should be defined, e.g., an acceptable amount of delay, and justified.

### 2.5.4. Original

The documents and the related materials should be original, and if it is not an original, it should be an exact copy. Also, the first record should be made by an appropriate person.

### 2.5.5. Accurate

The documentation should be accurate, consistent and a real representation of the facts. This includes whether the recorded information describes the conduct of the study without error, whether the conduct of the study conforms to the protocol, and the information about the people, who made the corrections and their specific timestamps [17].

It is quite essential to follow best practices to maintain proper DQ documentations, e.g., related references and materials should be properly maintained and updated. When a record needs to be modified, the previous data should not be obliterated. The person who makes the change should be identifiable and he should date the change and state the actual reason for the change. Also, the modifier should not whiteout the data or change the data without the knowledge whether the change is correct.

## 2.6. Measurement

The process of measurement is the act of ascertaining the size, amount, or degree of something. Measurements always involve comparisons, hence, measurements are the results of comparison. Most often, it includes a means to quantify the comparison.

People measure things all the time and our brains are hard-wired to understand unknown parts of our world in terms of things we know. It could be a complex process, as for those things we have not measured before, we often do not have a basis for comparison, the tools to execute the comparison, or the knowledge to evaluate the results. Measuring the quality of data, in particular in bio medical domain, is perceived as complex or difficult, because we often do not know what we can or should compare data against.

Measurement is objective and it is a repeatable way of characterizing the condition of the thing being measured. Also, this could be a beginning point for change, or a point for improvement of the thing that needs to be improved or can be considered as an indication of confirming that the improvement has taken place.

## 2.7. Assessment

Data quality assessment (DQA) is often subjective and is dependent on the evaluative task or the objective, which is particularly problematic, because clinical research datasets are often phenotype-specific, requiring a unique patient cohort, or may cater to a specific set of participating medical centers [21].

Although there are certain assurance checks that researchers can conduct to ensure data quality, this process is frequently time-consuming and cumbersome, and the results of these assessments may not be meaningful without a thorough understanding of the researcher's intended goal. Evaluating data missingness, distributions, and accepted values does not entirely paint a full picture of the quality of the dataset as described by the research objective, and manually evaluating quality in this fashion is a resource-intensive task. Also, there are no consistent evidence-based or community-driven metrics for assessing the quality of the research data. Study investigators frequently develop ad-hoc metrics that are specific to the study, and cannot be replicated [28], [30].

Like measurement, assessment requires comparison. But as with data quality measurement, data assessment, we do not always know what we are comparing data against. For example, how do we know what is wrong? What is an "error" by looking at a data sample? Assessment is not only about comparison, but it requires drawing in to conclusions and it depends on understanding the implications and how to act on them.

## 2.8. Data quality assessment

Ideally, data quality assessment (DQA) enables you to describe the condition of data in relation to particular expectations, requirements, or purposes in order to draw a conclusion about whether it is suitable for those expectations, requirements, or purposes. For example, organizations articulate expectations related to the expected condition or the quality of their data. Therefore, at the beginning of an assessment process, these expectations may not be known or fully understood. The assessment process includes uncovering and defining expectations.

Data assessment includes evaluation of how effectively data represents the objects, events, and concepts it is designed to represent. If you cannot understand how the data works, it will appear to be of poor quality. Data Assessment is usually conducted in relation to a set of quality dimensions that can be used to guide the process, esp. in the absence of clear expectations such as how complete the data is, how well it conforms to defined rules for validity, integrity, and consistency or how it adheres to defined expectations for presentation.

Assessment requires understanding, i.e. the concepts the data represents, the processes that created data, and the systems through which the data is created. Deliverables from an assessment include observations, i.e., what you see, implications, i.e., what it means and recommendations, i.e., what to do about it.

DQA is an iterative process. Often you will need to revisit earlier steps if you run into a problem in a later step. The Figure 1, shows one example of where iteration might occur. Step 4 in particular is one place you might run into a stopping point if your data does not meet the assumptions needed for the test you selected.

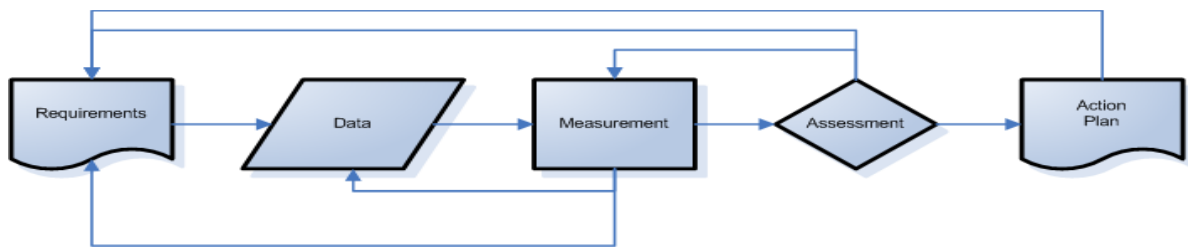

*Figure 1: DQA process*

# 3. Data Quality Activities

The data quality practitioner's primary function is to help an organization to improve and sustain the quality of their data, so that it gets optimal value from their available data.

Following are some of the key activities in order to improve and sustain the quality of the data;

- Defining / documenting quality requirements for data
- Measuring data to determine the degree to which data meets these requirements
- Identifying and remediating root causes of data quality issues
- Monitoring the quality of data in order to help sustain quality

- Partnering with business process owners and technology owners to improve the production, storage, and use of an organization's data
- Advocating for and modeling a culture committed to quality

Also, assessment of the condition of data and ongoing measurement of that condition are central to the purpose of the organization's data quality program.

## 3.1. Quality system components

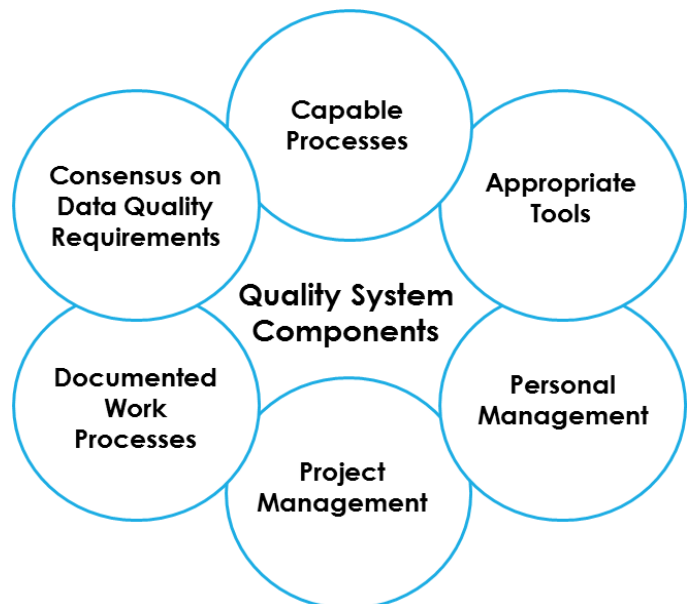

*Figure 2: Components of a data quality system. The environment in which data are collected and processed impacts data quality. Thus, achieving and controlling data quality usually requires action from entities in the broader environment [23]*

1. **Organizational consensus regarding the required level of data quality**, informed by an understanding of the cost of achieving it and the consequences of failing to achieve it.
2. **Appropriate tools** for supporting the collection and management of data. Although specialized devices and software are of themselves neither necessary nor sufficient for producing quality data, their presence is often perceived as representing rigor or important capabilities.
3. **Design of processes capable of assuring data quality**, likened to mass customization, in medical research, scientific differences in studies and circumstances of management by independent research groups drive variation in data collection and processing.
4. **Documented standard operating procedures (SOPs)** are required by FDA regulation and in most research contracts. The complete data collection and management process should be documented prior to the system development and its data collection processes.
5. **Personnel management** infrastructure; job descriptions, review of and feedback on employee performance, and procedures for managing performance.
6. **Project management** in medical research informatics begins with understanding the basic data-related requirements of a study, i.e., the data deliverables, associated costs, the necessary levels of quality, and the amount of time required or available. Project management also includes

14 / 19

planning to meet requirements as well as ongoing tracking, assessment, and reporting of status with respect to targets.

## 3.2.  Impact of DQ on research results

Unfortunately, a "one size fits all" data quality acceptance criterion is not possible for bio bank research, as its related statistical tests may vary in their robustness to data errors. The impact on the statistical test depends on the variable in which the errors occur and the extent of the errors. Further, the data that are of acceptable quality for one use may not be acceptable for another, i.e., the "fitness for use" aspect that is addressed earlier. Due to these reasons regulators cannot set a data quality minimum standard or an "error rate threshold."

The poor DQ may lead or facilitating condition that can lead to erroneous interpretations of the reality, which is misrepresented by the data. Consequently, it may lead to make wrong decisions and medical errors or may lead to adverse events, i.e., the patient is either harmed, damaged or even killed.

It is widely known that variances in data quality reduce the statistical power of the sample size. So, if the data variation is high, it requires more data to analyze, e.g. in medical domain, it requires more patients enrolled for the data collection, which consequently delay the clinical trial completion timelines. However, the real-life data is dirty; it is inconsistent, duplicated, inaccurate, and out of date. All these data defects contribute towards data variability, which lowers statistical power.

As shown in Fig. 3, lower data quality exhibit higher variability and lower confidence intervals compared to higher data quality, which yields higher confidence intervals. It is important to emphasize that study teams need to focus on improving data quality from all aspects; on biomedical perspective, this can include patient nonadherence, variability with coordinator mandated measurements, source data quality, and data collection methodologies. The ultimate benefit of improving data quality involves the notion that study teams will need to enroll less patients because of sufficient statistical power.

Further, caring about data consistency is also an important task, i.e., refers to the validity of the data source in which it is stored, and data integrity attributes towards the accuracy of the data that is stored within the data source.

To elaborate, from a data integrity standpoint, a coordinator may misinterpret and incorrectly record medical measurements on a paper CRF, e.g., without automated validations, a coordinator will not know they made an error until they input the data into EDC. From a consistency perspective, paper source introduces all kinds of risks. For example, if the coordinator loses the paper source, there is no validity towards the data in EDC, and paper source can be modified without validated tracking systems (i.e., erasing or rewording measurements and reproducing paper CRFs from memory). eSource is known to significantly improve both data consistency and integrity.

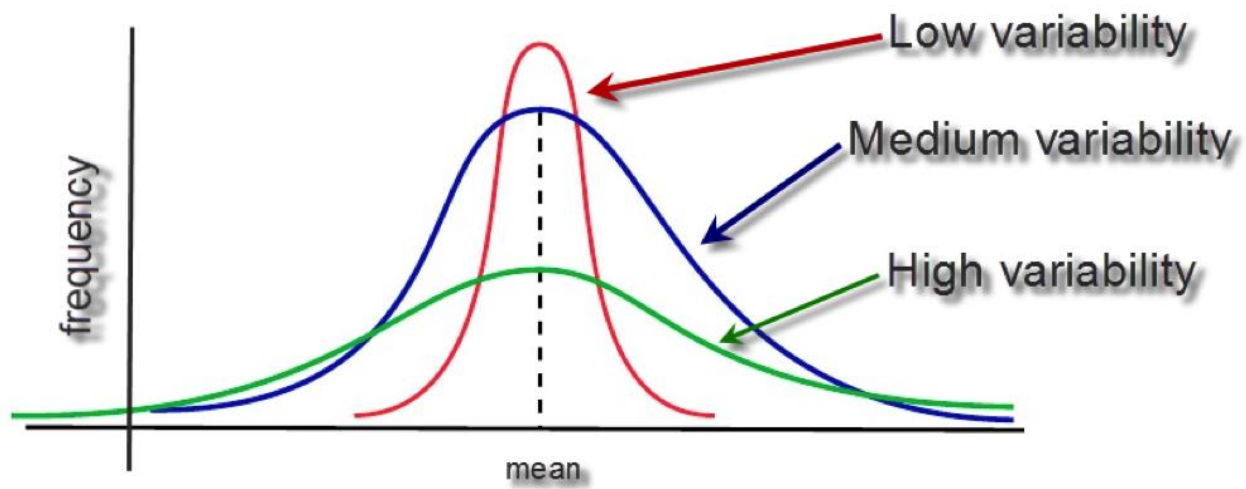

*Figure 3: Impact of data variability on statistical power [1]*

### 3.3. Common problems

This section will highlight some of the common problems that are related to the quality of the data and its related documentations.

Data needs to be interpreted and for interpretation it needs a context. The common problems for documenting data quality is that this context is not registered adequately. Main reasons for that is:
- unawareness of the respective context (e.g. which precision does a particular method of measurement have),
- lack of documentation of a particular context (e.g. it is known that person A is a trained pathologist, there is only 1 test to measure a certain property, etc.)
- evolution, relevant parameters of the context change. Old data are no longer interpretable correctly in the new context (cf. [14,15])

The main problem of knowing data quality stems from situations where data items lack of proper documentations, hence, missing dates, e.g., stored date, required temperatures, missing attributes, or subject identifiers, etc. Also, missing proper standards for recording the data items, e.g., no properly defined checklists and no proper data capturing and storing methods lead to inaccurate/erroneous transcription of values.

Also, practitioners are not having the habit of documenting the data assuming, the related information are generally known, e.g. if diagnosis by Prof. Cogito, or by a trained pathologist with more than 4 years of experience, as a habit such data are not properly documented assuming it is generally known. Similarly, if there is only 1 machinery in a lab to produce the data and the related results, such data origin information are not documented assuming it is generally known.

Further, the data related evolution or the changes are not continuously documented assuming such information are necessary. This generally happens due to the lack of knowledge about its intended use in the future.

# 4. Conclusion

Recent studies showed an alarming low level of reproducibility of medical studies. One of the major sources of irreproducibility is the lacking assessment and documentation of the quality of the material and data used in the studies. Data quality definition, measurement, and assessment are therefore of equal importance as the assessment of sample quality and have to be part of a holistic quality management in biobanks.

As most of the biobanks organized within bbmri.at are hospital based or disease based biobanks, they are mainly data brokers collecting data from various sources and providing these data for medical studies, the assessment and documentation of the quality of the data sources is essential.

Part of the data provided by the biobank is also data about the data quality, and this documentation of data quality is relevant for the search of samples as are quality indicators of the biological samples. Storing the provenance of the data also makes it easier for biobanks to include data produced elsewhere for example on basis of samples provided by the biobank in their offerings for researchers. Large biobanks may contain data and samples of quite varying quality. Therefore, it is important that data quality is not seen as an overarching characteristic of a whole biobank, but has to be described on the level of collections and samples. A quality characteristic of the whole biobank, nevertheless, is the quality of the whole data quality management system.